\documentclass[aip, jcp, preprint, notitlepage, superscriptaddress]{revtex4-1}

\usepackage[utf8x]{inputenc}
\usepackage[T1]{fontenc}
\usepackage{lmodern}
\usepackage{amsmath, amssymb}
\usepackage{mathrsfs}
\usepackage{braket}
\usepackage{booktabs}
\usepackage{graphicx}
\usepackage{array, dcolumn, makecell, multirow}
\usepackage{siunitx}
\usepackage[version = 3]{mhchem}
\usepackage[printonlyused, withpage]{acronym}
\usepackage{tikz}
\usepackage{paralist}
\usetikzlibrary{positioning}
\usetikzlibrary{calc}
\usetikzlibrary{arrows, automata, shadows, shapes}
\usepackage[colorlinks = true, linkcolor = blue, citecolor = blue, urlcolor = blue, bookmarks = true]{hyperref}
\usepackage[a4paper, left = 2cm, right = 2cm, top = 2cm, bottom = 2cm]{geometry}
\usepackage[normalem]{ulem}

\newcommand{\beq}{\begin{equation}}

\newcommand{\eeq}{\end{equation}}

\newcommand{\asop}{\ensuremath{\mathscr{A}}}
\newcommand{\mpcorr}{\ensuremath{\Delta \widetilde{E}^{(2)}}}

\begin{document}

\title{The nature of three-body interactions in DFT: exchange and polarization effects}
\author{Michał Hapka}
\email{hapka@tiger.chem.uw.edu.pl}
\affiliation{Faculty of Chemistry, University of Warsaw, ul. L. Pasteura 1, 02-093 Warsaw, Poland}
\author{Łukasz Rajchel}
\email{L.Rajchel@icm.edu.pl}	
\affiliation{Faculty of Chemistry, University of Duisburg-Essen, Universit{\"a}tsstraße 5, 45117 Essen, Germany}
\author{Marcin Modrzejewski}
\affiliation{Faculty of Chemistry, University of Warsaw, ul. L. Pasteura 1, 02-093 Warsaw, Poland}
\author{Rainer Sch\"{a}ffer}
\affiliation{Faculty of Chemistry, University of Duisburg-Essen, Universit{\"a}tsstraße 5, 45117 Essen, Germany}
\author{Grzegorz Chałasiński}
\affiliation{Faculty of Chemistry, University of Warsaw, ul. L. Pasteura 1, 02-093 Warsaw, Poland}
\author{Małgorzata M. Szczęśniak}
\affiliation{Department of Chemistry, Oakland University, Rochester, Michigan 48309-4477, United States}

\begin{abstract}

We propose a physically motivated decomposition of DFT 3-body nonadditive interaction energies into the exchange and density-deformation (polarization) components. 
The exchange component represents the effect of the Pauli exclusion in the wave function of the trimer and is found to be challenging for density functional approximations~(DFAs).
The remaining density-deformation nonadditivity is less dependent upon the DFAs.
Numerical demonstration is carried out for rare gas atom trimers, Ar$_2$-HX (X = F, Cl) complexes, and small hydrogen-bonded and van der Waals molecular systems. 
None of the tested semilocal, hybrid, and range-separated DFAs properly accounts for the nonadditive exchange in dispersion-bonded trimers. By contrast, for hydrogen-bonded systems range-separated hybrids achieve a qualitative agreement to within 20\% of the reference exchange energy. A reliable performance for all systems is obtained only when the monomers interact through the Hartree-Fock potential in the dispersion-free Pauli Blockade scheme.
Additionally, we identify the nonadditive second-order exchange-dispersion energy as an important but overlooked contribution in force-field-like dispersion corrections. 
Our results suggest that range-separated functionals do not include this component although semilocal and global hybrid DFAs appear to imitate it in the short range.
\end{abstract}

\maketitle

\section{Introduction}

The decomposition of the interaction energy into many-body contributions provides insights into the effects beyond pairwise contacts. Although 3- and higher-body components are significantly weaker than 2-body ones, the number of the former increases rapidly with the cluster size.~\cite{Szalewicz:2005} 
They cannot be neglected when predicting the structure and properties of clusters of water,\cite{Mas:2003b,Kamiya:2008,Gillan:12,Gillan:13a,Gillan:14,Erin:2016,Gillan:2016} clathrate hydrates,\cite{Deible:2014,Cox:2014} molecular crystals\cite{Beran:16} and condensed-phase systems.~\cite{Szalewicz:2005,Bukowski:2007}
Finally, the quantitative description of nonadditive effects is indispensable when resolving the high-resolution spectra of molecular trimers and larger aggregates.~\cite{Hutson1989,Cooper1993,Suhm:1995,Farrell:1996,Sperhac:96,Chalbie:00,Nauta:2001}

For more than a decade the applications of DFT to noncovalent interactions have focused on the development of dispersion-correction schemes accounting for the dispersion effects missing from hybrid and semilocal density functional approximations (DFAs).~\cite{Grimme:16}
As a result, the dispersion-inclusive DFT methods achieved statistical errors within a few tenths of a~kcal/mol with respect to \textit{ab initio} benchmark databases for noncovalent dimers.~\cite{Burns:2011,Podeszwa:12,Cohen:2012,Becke:14} 
Nevertheless, the studies of larger aggregates drew attention to the importance of many-body errors in DFT and to their origin.~\cite{Jordan:2010,Gillan:12,Gillan:13a,Ambrosetti:14,Gillan:14,Deible:2014} 

Recent work lists the following sources of DFA errors in binding energies:\cite{Erin:2016} a) errors in the intersystem exchange interaction, b) the delocalization error, and c) the monomer relaxation errors.
The error in the exchange interaction stems from the inability of the semilocal functional to describe the electron exchange in the tails of overlapping densities, i.e. regions of significant change of reduced density gradients $s = |\nabla \rho|/\rho^{4/3}$ upon formation of a noncovalent bond.~\cite{Lacks:93,Becke:09,Gillan:14}
In the language of perturbation theory this translates into the flawed treatment of the exchange terms arising from the Pauli exclusion principle.
The delocalization error, also known as many-electron self-interaction error, refers to the fact that semilocal DFAs yield electron densities which are too delocalized.~\cite{Cohen:2008,Cohen:2012} 
This is pronounced in systems with strong induction and electrostatic interactions, although one should not forget that all interaction energy components are affected.~\cite{Hapka:14}
Finally, the monomer relaxation error, which contributes to the binding energies of the cluster, concerns the ability of the functional to account for the energetics of bond stretching due to the formation of the complex. 
The inclusion of pair-wise dispersion corrections cancels out some of these errors.~\cite{Modrzejewski:14,Erin:2016}

Several studies have recognized the flaws in exchange functionals as a source of errors in 3- and higher-body terms.~\cite{Jordan:2010,Gillan:14,Rezac:15}
The most comprehensive test of DFAs encompasses a comparison of DFT and coupled-cluster total 3-body energies on the 3B-69 database of 23 molecular trimers in 69 configurations.~\cite{Rezac:15}
A common suggestion in the aforementioned works is that the errors arising from the incorrect many-body exchange may be more important than the missing many-body dispersion interactions. Clearly, this issue has to be addressed in order to improve the reliability of DFT predictions for large aggregates.

The conclusions in the existing literature (with only some exceptions -- see Ref.~\onlinecite{Jordan:2010}) are based on inference from the supermolecular computations of many-body terms which bundle together effects of a different physical origin and geometry-dependence. Unlike in the wave function methods, such as M{\o}ller-Plesset perturbation theory, where the many-body interaction energy components residing in each order are known,\cite{Chalbie:00} such insights are missing in the case of DFAs.

The performance of DFT for 3-body interactions can be understood given a physically meaningful decomposition of the interaction energy. To this end, we define the decomposition of the total 3-body interaction energy into the exchange and polarization components. Our aim is to reveal which nonadditive contributions to 3-body interaction energy are captured by different DFAs and how accurate this description is in DFT. This will be done by comparing the DFT exchange and polarization components with rigorously defined many-body terms from the perturbation expansion of the interaction energy. 

The assessment of the DFT’s ability to account for the exchange nonadditivity involves the following strategy. To extract the exchange part from a supermolecular DFT 3-body term we evaluate the interaction energy of the L{\"o}wdin-orthogonalized monomers which is the DFT equivalent of the Heitler-London exchange nonadditivity.~\cite{Cybulski:03} 
Next, we substitute the Hartree-Fock formula for the intermolecular part of the exchange potential to investigate how switching from the DFT to HF description changes the exchange nonadditivity. This is carried out with the aid of the dispersion-free Pauli Blockade (PBdf),\cite{Rajchel:10} the method which we generalized here to the many-body case.
Finally, nonadditive exchange energies extracted from both supermolecular DFT and PBdf are compared with the Heitler-London term, for which we provide benchmark data. This strategy serves as the first step toward revealing what 3-body contributions are captured in the supermolecular DFT interaction energies and how trustworthy their representation in specific DFT approximations is.

In the numerical secion we focus our attention on a range of challenging systems: from rare-gas trimers, to Ar$_2$-HX (X = F, Cl), to selected hydrogen and dispersion-bonded molecular trimers. 
Based on the presented energy decomposition scheme and 3-body symmetry-adapted perturbation theory (SAPT) analysis we draw conclusions on the performance of approximate DFT. 

\section{Theory}

\subsection{Energy decomposition}

The total 3-body nonadditive interaction energy is defined in terms of trimer, dimer and monomer total energies as: 
	\beq
        E_\mathrm{int} = E^\mathrm{ABC} - E^\mathrm{AB} - E^\mathrm{BC} - E^\mathrm{AC} + E^\mathrm{A} + E^\mathrm{B} + E^\mathrm{C}.
        \label{3ben}
	\eeq
Since our focus is solely on nonadditive interactions in trimers, we omit the~$[3, 3]$ label usually applied in the literature.
To analyze DFAs we decompose $E_{\rm int}$ into the nonadditive exchange and deformation contributions
\beq
E_{\rm int} = E_\text{nadd-ex} + E_\mathrm{def}.
\label{endecomp}
\eeq

The nonadditive exchange captures the energetic effect of applying the Pauli exclusion in the Kohn-Sham wave function upon formation of the trimer out of noninteracting monomers.
The computation of $E_\text{nadd-ex}$ for an arbitrary DFA consists of a noniterative trimer and dimers energy evaluation from the L{\"o}wdin-orthogonalized\cite{Lowdin:50} orbitals of the isolated monomers.~\cite{Cybulski:03} $E_{\rm int}$ is computed directly from Eq.~\eqref{3ben}.
The deformation energy corresponds to the energy lowering upon mutual self-consistent polarization of the monomers restrained by the Pauli exclusion principle and is computed as 
	\beq
		E_{\rm def} = E_{\rm int} - E_{\rm nadd-ex}.
	\eeq

An analogous energy decomposition for dimers was first proposed by Cybulski and Seversen.~\cite{Cybulski:03,Cybulski:05}
$E_{\rm nadd-ex}$ is the DFT version of the nonadditive Heitler-London interaction energy, a well-established concept in the theory of intermolecular interactions:\cite{Lowdin:56,Cybulski:03,Cybulski:05}
	\beq
		E^\mathrm{HL}_{\rm int} = \frac{\Braket{ \asop \Phi_{0} | H | \asop \Phi_{0} }}{\Braket{ \asop \Phi_{0} | \asop \Phi_{0} }} - \sum_{\mu = {\rm A, B, C}} \frac{\braket{\Phi^{\mu}_0|H_{\mu}|\Phi^{\mu}_0}}{\braket{\Phi^{\mu}_0|\Phi^{\mu}_0}} - E_{\rm int}^{\rm HL}({\rm AB}) - E_{\rm int}^{\rm HL}({\rm BC}) - E_{\rm int}^{\rm HL}({\rm CA}),
\label{hldef}
	\eeq
where $\Phi_{0} =\Phi^{\rm A}_0\Phi^{\rm B}_0\Phi^{\rm C}_0$, $\Phi_0^{\mu}$ is the wave function of the monomer $\mu$, {\asop} is the antisymmetrizer which exchanges electrons between monomers, and $E_{\rm int}^{\rm HL}$(XY) is the Heitler-London interaction energy of the XY dimer. 

\subsubsection{Nonadditive exchange term}

The energy partitioning introduced in Eq.~\eqref{endecomp} is physically appealing due to the fact that benchmark values for the $E_{\rm nadd-ex}$ term can be calculated directly from symmetry-adapted perturbation theory (SAPT). The reference for $E_{\rm nadd-ex}$, denoted $E^{\rm HL(KS)}_{\rm int}$, is computed as the Heitler-London interaction energy with Kohn-Sham determinants.

In order to use the existing 2-body and 3-body codes for SAPT based on Kohn-Sham description of monomers (DFT-SAPT) we compute this term as
\beq
E^{\rm HL(KS)}_{\rm int} = E^{(1)}_{\rm exch,SAPT} + \Delta_\mathrm{M},
\label{ehlks}
\eeq
where the first-order DFT-SAPT nonadditive exchange energy is 
	\beq
		E_\mathrm{\rm exch,SAPT}^{(1)} = \frac{\Braket{ \Phi^{\rm KS}_0 | \asop V | \Phi^{\rm KS}_0 }}{\Braket{ \Phi^{\rm KS}_0 | \asop | \Phi^{\rm KS}_0 }} - E^{(1)}_{\rm exch,SAPT}({\rm AB}) - E^{(1)}_{\rm exch,SAPT}({\rm BC}) - E^{(1)}_{\rm exch,SAPT}({\rm CA}),
	\eeq
and $\Delta_\mathrm{M}$ denotes the Murrell delta term,\cite{JBP:76} also referred to as the zeroth-order exchange energy. (See the Supplementary Information (SI) for explicit formulas for the computation of $\Delta_{\rm M}$.)  

The reference $E^{\rm HL(KS)}_{\rm int}$ value depends only weakly on the underlying DFA, provided that an asymptotic correction\cite{Cencek:13} or a tuned range-separated functional\cite{Baer:10,Modrzej:13} is applied (see Table~S1 in SI for the sensitivity test of $E^{\rm HL(KS)}_{\rm int}$). The correct asymptotic behavior of the exchange-correlation potential yields reliable monomer densities', a well-known prerequisite for DFT-SAPT calculations.~\cite{Hesselmann:02}

Based on the DFT-SAPT performance for the two-body first-order exchange energy,~\cite{Korona:13} we expect that the 3-body $E^{(1)}_{\rm exch,SAPT}$ term is most accurately reproduced when monomers are described with asymptotically-corrected hybrid functionals. Consequently, we will use $E^{\rm HL(KS)}_{\rm int}$ based on PBE0\cite{Perdew:96,Adamo:99} functional with the GRAC\cite{Gruning:2001} asymptotic correction as a reference for the nonadditive exchange.

Depending on the quality of the exchange-correlation functional, $E_{\rm nadd-ex}$ should approach the reference  $E^{\rm HL(KS)}_{\rm int}$ value. While both $E_{\rm nadd-ex}$ and $E^{\rm HL(KS)}_{\rm int}$ are computed with approximate DFT methods, the latter replaces the exchange-correlation functional in the intermonomer region with the Heitler-London exchange. The description of the intramonomer correlation enters in both quantities through the use of Kohn-Sham orbitals.~\cite{Podeszwa07}

It should be stressed that the major part of the exact nonadditive exchange energy is captured already at the Hartree-Fock level of theory.~\cite{Podeszwa07} 
In particular, the $E_{\rm nadd-ex}$ contribution which is included in the supermolecular 3-body Hartree-Fock interaction energy
is equal to the $E^\mathrm{HL}_{\rm int}$ formula calculated with Hartree-Fock determinants.
Obviously, Hartree-Fock misses the intramonomer correlation contribution to $E_{\rm nadd-ex}$ which in the M{\o}ller-Plesset perturbation theory enters at the MP2 and higher levels.~\cite{Chalbie:00} 

\subsubsection{Deformation term}
Several comments regarding the $E_{\rm def}$ component should be made. At the Hartree-Fock level of theory the deformation contribution collects second- and higher-order induction energy components together with their exchange counterparts while the lowest-order intramonomer contributions to those terms enter at the MP2 level.~\cite{Chalbie:00} We expect that the nonadditive many-body induction terms, including the intramonomer correlation contributions, should be recovered quantitatively by the exisiting DFAs and their description should systematically improve upon the minimization of the delocalization error.~\cite{Rezac:15,Erin:2016} On the other hand, the third-order induction-dispersion energy (along with its exchange counterpart) and second-order exchange-dispersion energy -- both included at the MP2 level -- are most problably not accounted for due to the semilocal nature of DFAs. We will investigate this problem in more detail in Section III.B.

A well-known component of the many-body nonadditive interaction energy is the third-order dispersion energy, $E^{(3)}_{\rm disp}$, which in the M{\o}ller-Plesset theory first appears at the MP3 level. Much attention has been devoted to the proper inclusion of this term in DFT.~\cite{Grimme:16,Hermann:17} We address the importance of $E^{(3)}_{\rm disp}$ relative to the remaining second- and third-order terms in Sections III.B and III.C. 

\subsection{Dispersion-free Pauli Blockade}

\begin{figure}[htbp]
\includegraphics[scale=0.9]{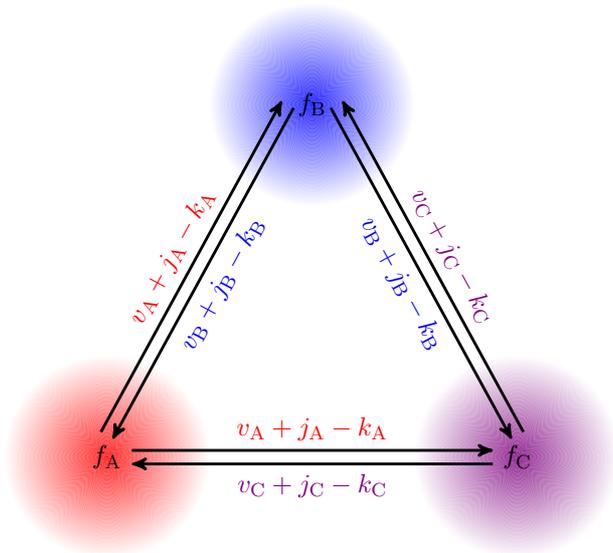}
		\caption{3-body dispersion-free Pauli Blockade scheme. The monomers are described with the full KS operator~$f_{\mu}$; they interact through the HF operators~$v_{\mu} + j_{\mu} - k_{\mu}$.}
		\label{fig:pb_physics}
\end{figure}

To expose the DFA errors in the description of intermonomer regions, 
we introduce a scheme in which the Kohn-Sham exchange in the intermonomer region is replaced by the exact Hartree-Fock expression (Figure~\ref{fig:pb_physics}). This approach eliminates the dispersion component of the interaction energy and is referred to as dispersion-free Pauli Blockade.~\cite{Rajchel:10}
In this section we briefly introduce the idea behind many-body PBdf and comment on the contents of the nonadditive 3-body PBdf interaction energy.

In the PBdf formalism the Kohn-Sham equation for the orbitals of monomer $\mu$ in a system composed of $L$~weakly interacting monomers is
\beq
		\Bigg( f_\mu(\mathbf{r}) + \sum_{\nu \neq \mu}^L \Big( v_{\nu}(\mathbf{r}) +
			j_{\nu}(\mathbf{r}) - k_{\nu}(\mathbf{r}) \Big)
		\Bigg) \varphi_\mu^i(\mathbf{r}) =
		\epsilon_\mu^i \varphi_\mu^i(\mathbf{r}),
		\label{pbdf:eqs}
\eeq
where $f_{\mu}(\mathbf{r})$ is the Kohn-Sham operator of the $\mu$ monomer, $v_{\nu}(\mathbf{r})$ describes nuclei-electron attraction,
$j_{\nu}(\mathbf{r})$ is the Coulomb electron-electron repulsion, and $k_{\nu}(\mathbf{r})$ is an exact HF-like exchange operator. Note that $f_{\mu}(\mathbf{r})$ corresponds to the unaltered Kohn-Sham operator for a given DFT approximation. 

Eq.~\eqref{pbdf:eqs} is solved in an iterative fashion keeping the monomer orbitals orthogonal at all times. This may be achieved either with the use of the penalty function,\cite{Rajchel:10cpl} or the exponential ansatz of orbital rotation.~\cite{Modrzej:12} The latter scheme is more efficient and numerically stable, and therefore has been used in the current implementation of the PBdf method. The iterative process stars from the L{\"o}wdin-orthogonalized\cite{Lowdin:50} orbitals of the isolated monomers.

The converged orbitals are used to compute the dispersion-free energy of the complex:
\beq
\begin{split}
E^\mathrm{dfree}[\tilde{\rho}_1, \tilde{\rho}_2, \ldots, \tilde{\rho}_L] & =
	\sum_{\mu = 1}^L E_\mu[\tilde{\rho}_\mu] \\  
 & + \sum_{\mu = 1}^{L - 1} \sum_{\nu = \mu + 1}^L \Big(
 E_\mathrm{elst}[\tilde{\rho}_\mu,  \tilde{\rho}_\nu] +
 E_\mathrm{exch}[\tilde{\rho}_\mu,  \tilde{\rho}_\nu] \Big)
\end{split}
\label{pbdf:enfun}
\eeq
where~$E_\mu[\tilde{\rho}_\mu]$ is the monomer energy functional, $\tilde{\rho}_{\mu}$ denotes the density of the monomer $\mu$ obtained from the converged orthogonalized orbitals, the electrostatic interaction term reads
	\beq
		\begin{split}
			E_\mathrm{elst}[\tilde{\rho}_\mu,  \tilde{\rho}_\nu] & =
				\int_{\mathbb{R}^3} v_\nu(\mathbf{r}) \tilde{\rho}_\mu(\mathbf{r}) \, \mathrm{d}^3\mathbf{r} +
				\int_{\mathbb{R}^3} v_\mu(\mathbf{r}) \tilde{\rho}_\nu(\mathbf{r}) \, \mathrm{d}^3\mathbf{r}  \\ 
			 & + \int_{\mathbb{R}^3} \int_{\mathbb{R}^3} \frac{\tilde{\rho}_\mu(\mathbf{r}_1) \tilde{\rho}_\nu(\mathbf{r}_1)}{r_{12}}  \mathrm{d}^3\mathbf{r}_1 \, \mathrm{d}^3\mathbf{r}_2 + W_{\mu \nu},
		\end{split}
	\eeq
with $W_{\mu \nu}$ denoting the nuclear-nuclear repulsion energy, and $E_\mathrm{exch}$ is given by the HF formula.

The PBdf energy formula, Eq.~\eqref{pbdf:enfun}, is used to compute the dispersion-free total nonadditive 3-body interaction energy,~$E_{\rm int}^{\rm dfree}$, according to Eq.~\eqref{3ben}, which requires converging the orbitals in Pauli Blockade equations for the trimer and for the dimers. The energy partitioning of Eq.~\eqref{endecomp} yields dispersion-free counterparts of nonadditive exchange and deformation energy contributions:
\beq
E_{\rm int}^{\rm dfree} = E_{\rm nadd-ex}^{\rm dfree} + E_{\rm def}^{\rm dfree},
\eeq 
where the $E_{\rm nadd-ex}^{\rm dfree}$ term is obtained in the same manner as $E_{\rm nadd-ex}$ for the full DFA functional, i.e. employing orbitals of the isolated monomers which are L{\"o}wdin-orthogonalized but otherwise kept unperturbed. As will be shown, because of the enforcement of the exact exchange $E_{\rm nadd-ex}^{\rm dfree}$ varies much less than $E_{\rm nadd-ex}$ among different DFAs.

While the prescription for the PBdf energy is based on a simple model, it provides a useful diagnostic to gain insight into the source of qualitative errors in DFT exchange.
A detailed derivation of the Pauli Blockade and dispersion-free Pauli Blockade schemes may be found in SI and in Refs.~\onlinecite{Rajchel:10cpl,Rajchel:10}.

\subsection{Computational details}
The d5z basis of Ref.~\onlinecite{Mitek:2008} was chosen for He. The argon trimer, as well as Ar$_2$-HX (X = F, Cl) systems were studied with the aug-cc-pVQZ basis of Dunning.~\cite{Dunning:89} The molecular trimers from the 3B-69 basis\cite{Rezac:15} and Ref.\onlinecite{Erin:2016} were studied with the aug-cc-pVTZ basis. All 3-body energies were counterpoise-corrected.~\cite{Boys:1970} 
The MVS\cite{Sun:15} and SCAN\cite{Sun2015b} calculations were done on a large grid of 300 radial and 1202 angular points of the Lebedev grid.
The IP-optimized values of the range-separation parameter $\omega$ for the $\omega$PBE\cite{Henderson:2008} functional (also known as LRC-$\omega$PBE), $\omega$ = 0.55 for Ar and $\omega$ = 1.0247 for He, were taken from Ref.~\onlinecite{Lao:14}. For molecular trimers studied in Sec.~III.C  $\omega$PBE was used with $\omega = 0.5$, as recommended in Ref.~\onlinecite{Erin:2016}.
The LC-PBETPSS functional\cite{Modrzejewski:2016} represents the class of meta-GGA range-separated hybrids.
All LC-PBETPSS calculations functional were performed with the default $\omega = 0.35$ value. The DFT-SAPT calculations employed the PBE0 functional\cite{Perdew:96,Adamo:99} with the GRAC\cite{Gruning:2001} asymptotic correction. 
The first-order nonadditive exchange and second-order nonadditive exchange-dispersion SAPT contributions were obtained without overlap expansion, i.e., they are correct to all orders of the intermonomer overlap.~\cite{Rainer:2012,Rainer:2013,Jansen:2014,Rainer:2016,Rainer:tbp} Additionally, for molecular trimers analyzed in Sec.~III.C we calculated second-order exchange-dispersion energies at the uncoupled Hartree-Fock level ($E^{(2;0)}_{\rm exch-disp}(S^2+S^3)$) with the SAPT2012 code.~\cite{sapt12,Lotrich:3b97,Lotrich:00,Podeszwa07}
All 3-body PBdf and DFT-SAPT calculations were done in the developer's version of \textsc{Molpro}.\cite{molpro:12} Calculation with the $\omega$B97XD3\cite{Chai:2013} functional were done in the \textsc{Orca} program.~\cite{orca12}

\section{Results}

\subsection{Rare-gas atom trimers}

The failure of supermolecular DFT in predicting 3-body interactions in rare gas trimers was studied in the past.~\cite{Tka:08,Gillan:14}
The wrong description of regions with high reduced density gradients translates into the flawed nonadditive exchange interaction energies in the cluster.~\cite{Jordan:2010,Gillan:14,Erin:2016}
This assessment of the DFT performance is evidenced in Table~\ref{decomp}, where the total 3-body interaction energies seriously deviate from both Hartree-Fock and coupled-cluster values. 
Because the tested functionals do not include any dispersion correction it is reasonable to expect the DFT energies to be close to the Hartree-Fock ones. 

\begin{table}[h!]
\caption{3-body nonadditive exchange and total supermolecular nonadditive interaction energies ($\mu$Hartree) for helium and argon trimers at equilateral triangle geometries. $E^{\rm HL(KS)}_{\rm int}$ should be regarded as the reference for $E_{\rm nadd-ex}$.}
\begin{tabular}{l *{4}{S}}
\toprule
\multirow{2}{*}{method}  & \multicolumn{2}{c}{\ce{He3} ($R = \SI{5.6}{\bohr}$}) & \multicolumn{2}{c}{\ce{Ar3} ($R = \SI{7.0}{\bohr}$)} \\
	\cline{2-5}
       & \multicolumn{1}{c}{$E_{\rm nadd-ex}$} & \multicolumn{1}{c}{$E_{\rm int}$} & \multicolumn{1}{c}{$E_{\rm nadd-ex}$} & \multicolumn{1}{c}{$E_{\rm int}$} \\ \midrule
RHF    &  -0.778 & -0.869 & -14.35 & -15.75 \\
CCSD(T)& \textemdash & -0.315 & \textemdash & 14.84 \\
$E^{\rm HL(KS)}_{\rm int}$ & -0.936 & \textemdash & -17.51 & \textemdash \\
BLYP   & -15.10  & -14.71 & -46.85 & -39.17 \\
B3LYP  & -8.423  & -8.048 & -16.83 & -12.13 \\
PBE    & 26.68   & 27.31  & 150.6  &  158.1\\
PBE0   & 15.35   & 15.74  & 89.80  & 92.17 \\  
$\omega$PBE & -2.261 & -2.360 & -41.55 & -46.43\\
\bottomrule
\end{tabular}
\label{decomp}
\end{table}

For the comparison of DFAs we chose two functionals of the GGA rung which feature different asymptotic behavior of the exchange enhancement factor $F_x(s)$: BLYP\cite{b:88} and PBE\cite{pbe:96}, as well as their hybrid counterparts: B3LYP\cite{Becke:1993} and PBE0. In addition, the optimally-tuned $\omega$PBE functional represents the class of range-separated functionals. 

The functionals characterized by rapid increase of $F_x(s)$ (BLYP and its hybrids) predict much too attractive nonadditive exchange energy while the slow increase of $F_x(s)$ in PBE-based functionals results in much too repulsive $E_{\rm nadd-ex}$ (Table~\ref{decomp}). 
It is worthwhile to note that the similar behavior of the 3-body exchange terms in the B88- and PBE-based functionals was recognized before in the analysis of water hexamers.~\cite{Jordan:2010} 

Two factors determine the accuracy of the DFT nonadditive exchange energy: the quality of density in the region of large reduced density gradient and the approximations in the exchange-correlation energy formula.~\cite{Gillan:14,Erin:2016}
The progression of results from semilocal PBE, to a global hybrid PBE0, to range-separated $\omega$PBE functional shows that including a fraction of Hartree-Fock exchange is mandatory and the long range corrected form of the functional brings the biggest improvement (Table~\ref{decomp}). For the PBE-based functionals the long range corrected form of the functional is imperative to account even for the correct sign of $E_{\rm nadd-ex}$. 
Next, the effect of the HOMO orbital, which determines the tail of the density, is also important.
For example, for He$_3$ the $\omega$PBE functional with the default $\omega$ parameter corresponds to $E_{\rm nadd-ex} = -3.9$~$\mu$Hartree, whereas the variant of this functional which enforces Koopmans' theorem corresponds to $E_{\rm nadd-ex} = -2.3$ $\mu$Hartree. Note that the sign of $E_{\rm nadd-ex}$ is correct in B88-based functionals but the inclusion of the exact exchange results in further substantial improvement.

The Heitler-London exchange energy can be reproduced accurately provided that the interaction between the monomers is described fully at the Hartree-Fock level. 
This can be achieved by means of PBdf, where the monomers interact with the exact Hartree-Fock potentials but the intramonomer contributions to the interaction energy are described at the DFT level.
Applying the $\omega$PBE functional in the PBdf framework reduces the error in the Heitler-London energy by an order of magnitude with respect to $\omega$PBE for He$_3$ and Ar$_3$ (Table~\ref{rgpbdf}). 
Somewhat surprisingly, the remaining discrepancy between PBdf and $\omega$PBE suggests 
that the full Hartree-Fock exchange at long range in the range-separated hybrid is not sufficient to capture the effect of the nonadditive exchange. The reference values which PBdf should approach, provided that the underlying functional accurately describes intramonomer correlation, are the 3-body interaction energies at the MP2 level minus the second-order exchange-dispersion term. We reiterate that while MP2 does not include the three-body dispersion energy, it contains the uncoupled second-order exchange-dispersion.~\cite{chalbie:90,Chalbie:00} The latter was shown to be important in rare-gases.~\cite{Lotrich:97three,Lotrich:97,Lotrich:1998,Podeszwa07} (The induction correlation effects are also present but for rare gas trimers are negligible.)

\begin{table}[h!]
\caption{3-body interaction energy components and total interaction energies ($\mu$Hartree). PBdf results were obtained with $\omega$PBE. $E_{\rm exch-disp,KS}^{(2)}$ denotes the coupled Kohn-Sham results.}
\begin{tabular}{*{6}{S}}
\toprule
\multicolumn{1}{c}{$R$} & \multicolumn{1}{c}{$E^\mathrm{HL(KS)}_\mathrm{int}$} & \multicolumn{1}{c}{$E_\mathrm{nadd-ex}^{\mathrm{dfree}}$} & \multicolumn{1}{c}{$E_\mathrm{int}^{\mathrm{dfree}}$} & \multicolumn{1}{c}{$E_{\rm exch-disp,KS}^{(2)}$} & \multicolumn{1}{c}{$E_\mathrm{int}^\mathrm{MP2}$} \\ \toprule
\multicolumn{6}{c}{\ce{He$_3$}} \\ \midrule
4.0	& -204.9 & -195.3 & -205.5 & 26.38 & -183.6 \\
5.0	& -6.937 & -6.796 & -7.454 & 1.594 & -6.016 \\
5.6	& -0.849 & -0.847 & -0.942 & 0.279 & -0.688 \\
6.0	& -0.203 & -0.207 & -0.232 & 0.086 & -0.154 \\ 
\multicolumn{6}{c}{\ce{Ar$_3$}} \\ \midrule
6.0	& -209.4 & -210.8 & -228.2 & 148.3 & -123.8 \\
7.0	& -14.49 & -14.87 & -16.28 & 18.12 & -2.230 \\
7.5	& -3.688 & -3.799 & -4.165 & 6.020 &  0.635 \\
8.0	& -0.913 & -0.951 & -1.041 & 1.972 & 0.546 \\
\bottomrule
\end{tabular}
\label{rgpbdf}
\end{table}

\subsection{Ar$_2$-HX (X=F,Cl) }
The paradigm trimers for studies of nonadditive forces, Ar$_2$-HF and Ar$_2$-HCl, were the first systems for which it was possible to extract the 3-body potential based on the microwave\cite{Gutowsky1987,Klots1987} and far-infrared\cite{Elrod1991a,Elrod1993} spectroscopic data, as done in the works of Hutson \textit{et al.}\cite{Hutson1989} and Cooper and Hutson.~\cite{Cooper1993} The nature of nonadditive interactions in Ar$_2$-HX was later thoroughly studied in a series of papers combining supermolecular MP$n$ calculations and direct calculations of nonadditive energy components based on perturbation theory.\cite{Szczesniak:93,Cybulski:94,chalbie:97,Lotrich:1998,Moszynski:1998}
3-body interactions manifest themselves most clearly in the regions of in- and out-of plane rotations of HF and HCl. 

As shown in Refs.~\onlinecite{Szczesniak:93} and \onlinecite{Cybulski:94}, the dominant contribution to the anisotropy in Ar$_2$-HX comes from the exchange nonadditivity, yet the induction and dispersion components cannot be neglected. 
The leading contribution to the third-order induction comes from the moments induced on two argon atoms by the field of HX.
The exchange and induction nonadditivities are more pronounced in Ar$_2$-HF as the distance between HF and the center of mass of Ar$_2$ is considerably smaller than in Ar$_2$-HCl, and the HF dipole is larger than that of HCl.

\begin{figure}
\includegraphics[scale=0.6]{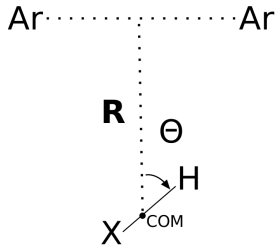}
\caption{Definition of geometrical parameters in Ar$_2$-HX.  
$\Theta$ is the angle between the $R$ vector and the HX axis. Ar$_2$ and HX are in one plane.} 
\label{fig:geom}
\end{figure}

In the present work we focus only on the in-plane rotations of HX, where $\Theta$ is varied from \ang{0} (the global minimum) to \ang{180} (Figure~\ref{fig:geom}). $\Theta =$ \ang{0} corresponds to H pointing toward the center of the triangle. The bond lengths for Ar$_2$HCl are:\cite{Szczesniak:93,Cybulski:94} $r_{\rm HCl}$ = \SI{1.275}{\angstrom}, $r_{\rm Ar-Ar}$ = \SI{3.861}{\angstrom} and $R = \SI{3.4736}{\angstrom}$. The values for Ar$_2$HF are: $r_{\rm HF}$ = \SI{0.917}{\angstrom}, $r_{\rm Ar-Ar}$ = \SI{3.826}{\angstrom} and $R = \SI{2.9798}{\angstrom}$.

\begin{figure}
\includegraphics[scale=0.65]{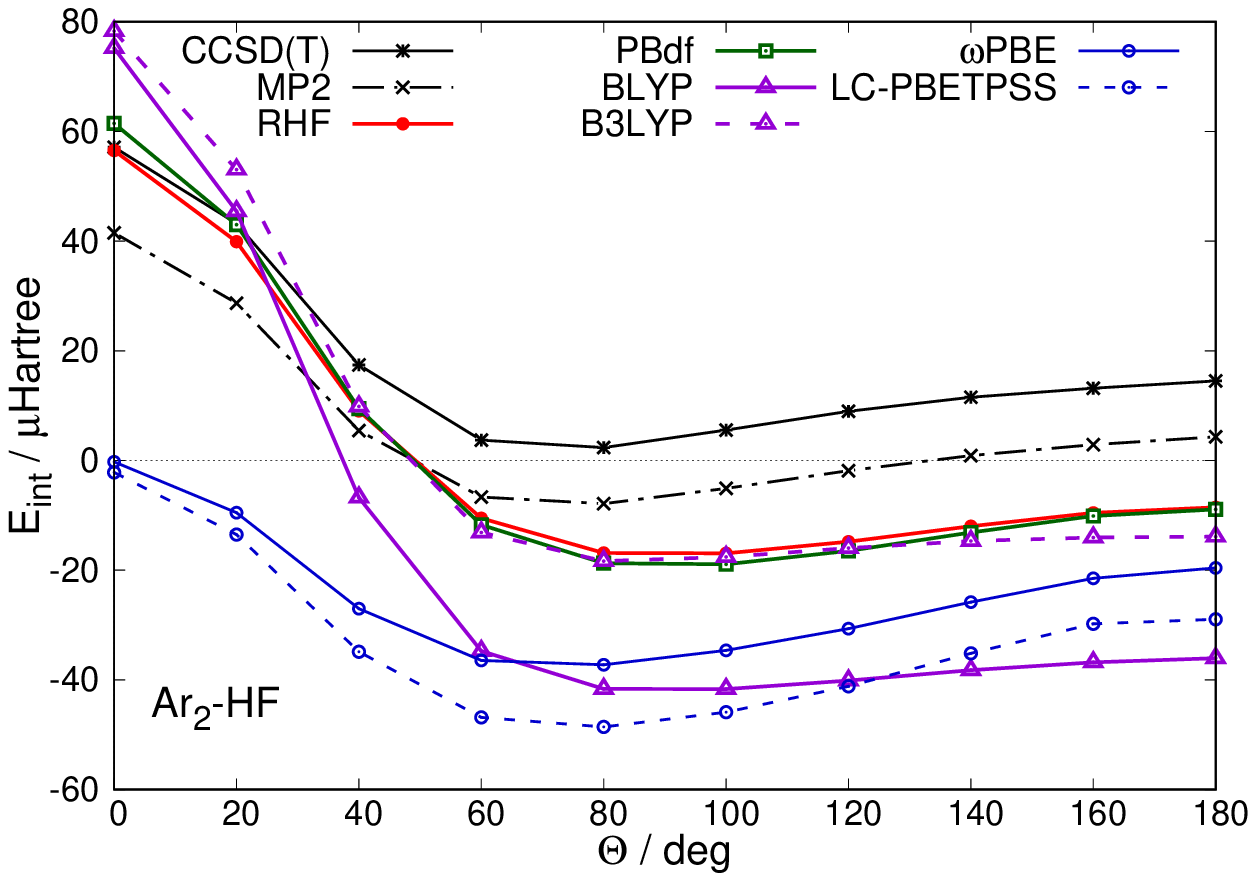}
\includegraphics[scale=0.65]{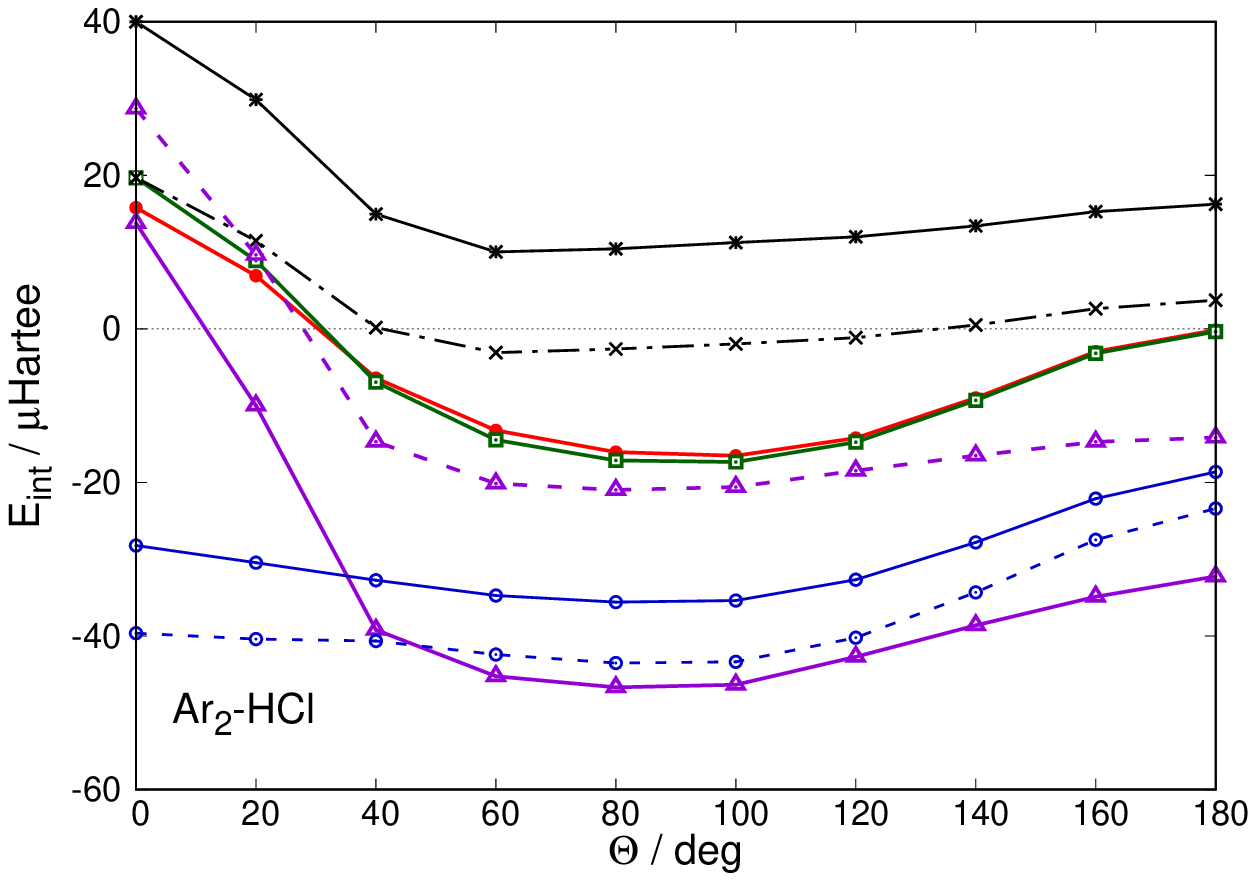}
\caption{3-body nonadditive interaction energy for the in-plane bend of the Ar$_2$-HF and Ar$_2$-HCl trimers. To avoid distorting the scale of the plot, we show the PBE and PBE0 results in SI. PBdf results were obtained with the PBE0 functional. }
\label{fig1:ar2hx}
\end{figure}

The performance of DFT methods for the total 3-body interaction energies of Ar$_2$-HX is qualitatively the same as in the case of noble gas trimers. The BLYP and B3LYP functionals are in better agreement with both Hartree-Fock and MP2 results than the PBE-based functionals (Figure~\ref{fig1:ar2hx}, the results for PBE and PBE0 are given in Figure~S1, SI). The behavior of the PBE-based approximations improves upon introducing range separation of the exchange functional. 
Finally, PBdf and Hartree-Fock, which do not account for the second-order exchange dispersion and induction-dispersion effects, closely follow each other for Ar$_2$-HX.

\begin{figure}
\begin{tabular}{l r}
\includegraphics[scale=0.65]{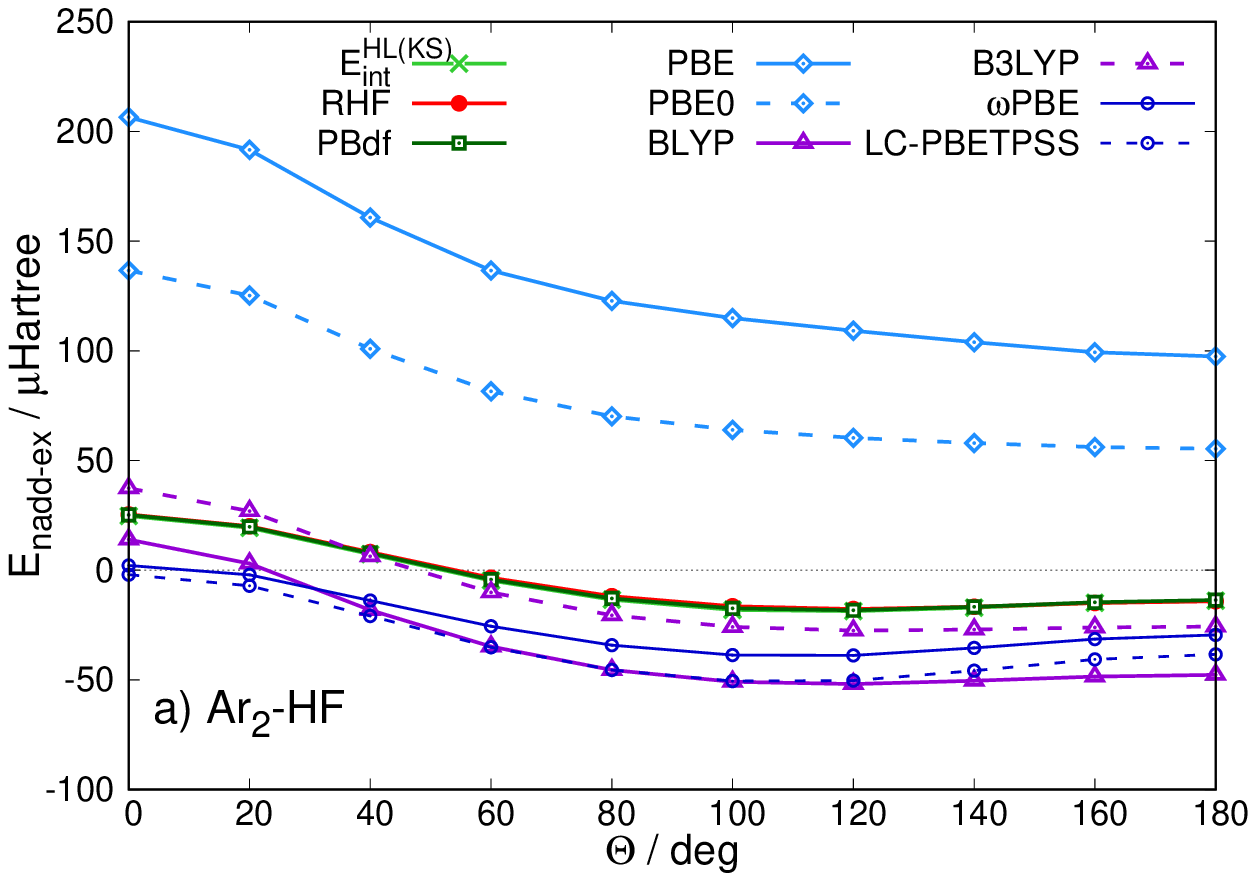} & \includegraphics[scale=0.65]{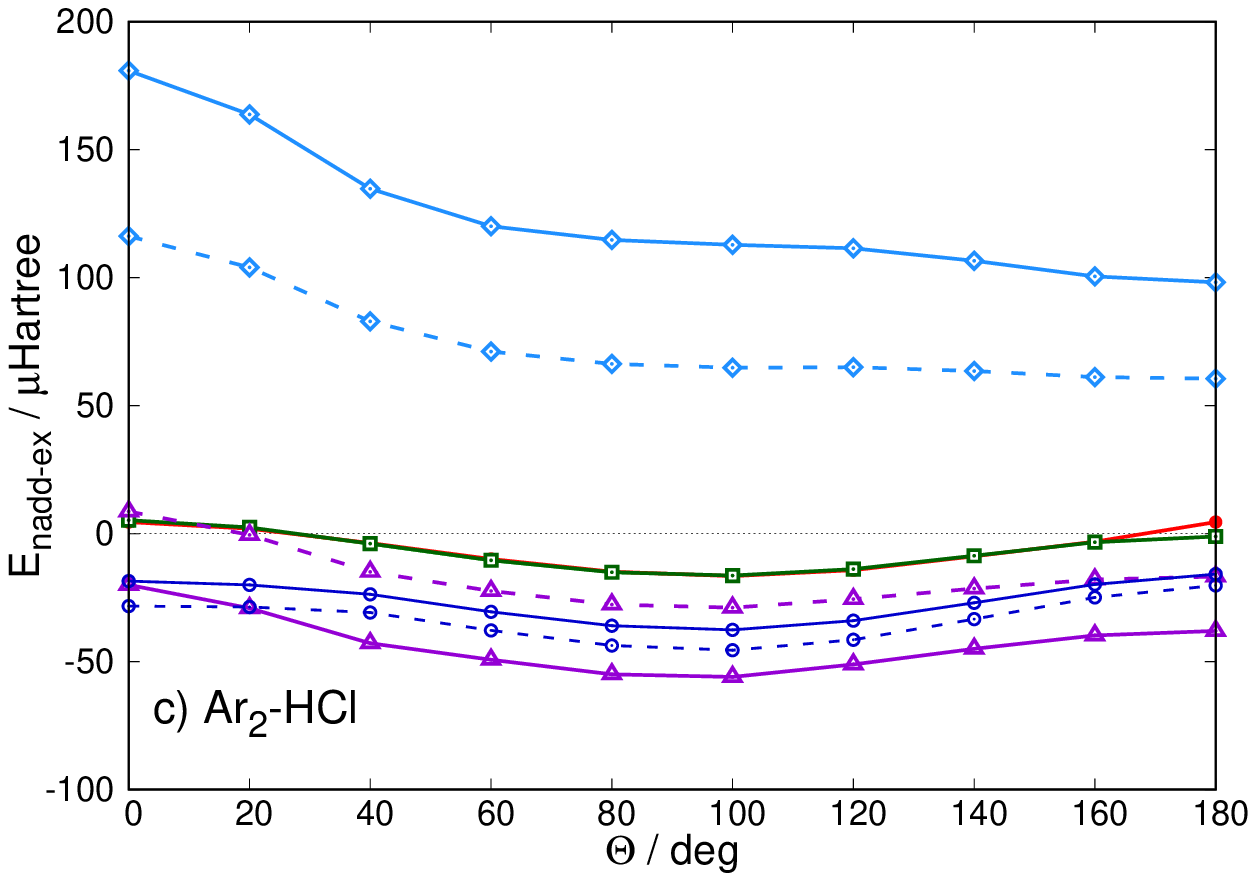} \\
\includegraphics[scale=0.65]{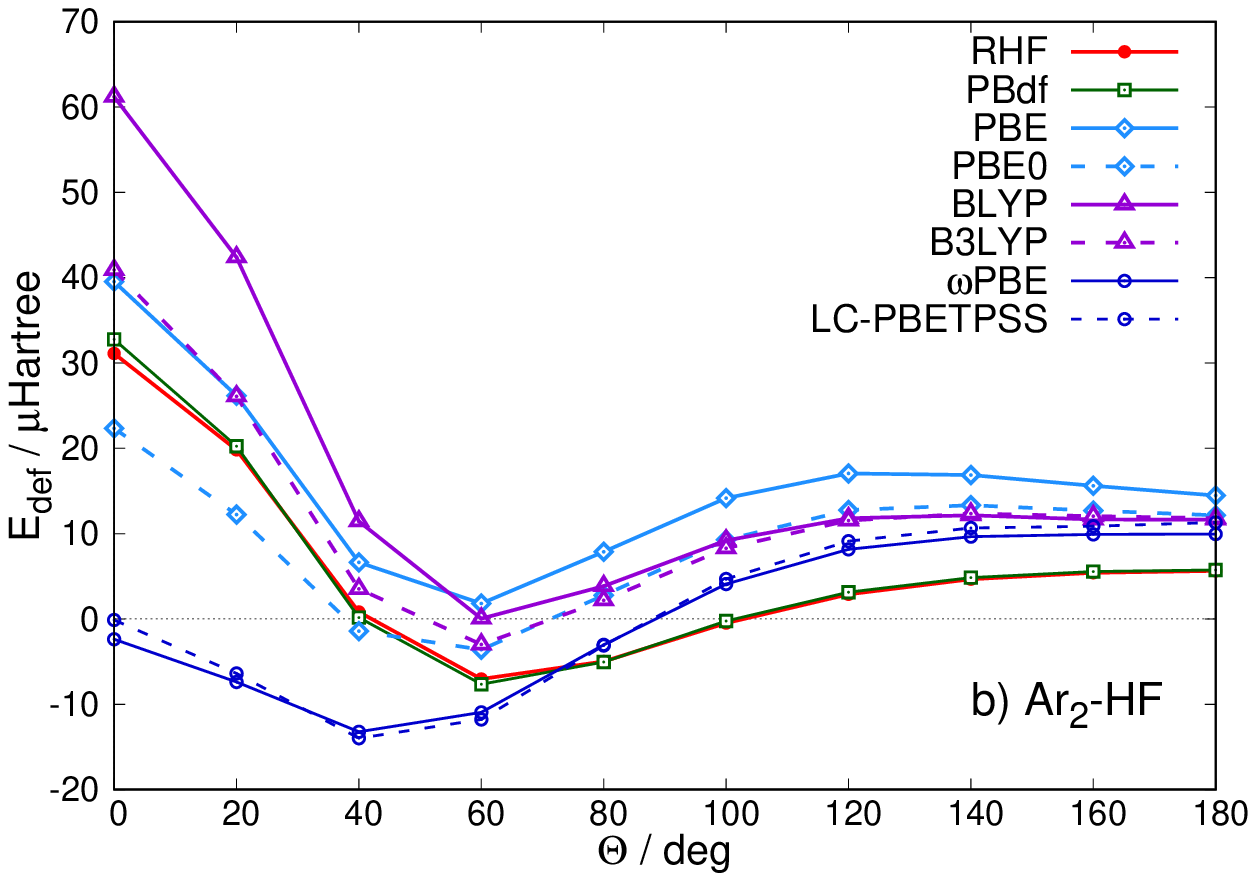} & \includegraphics[scale=0.65]{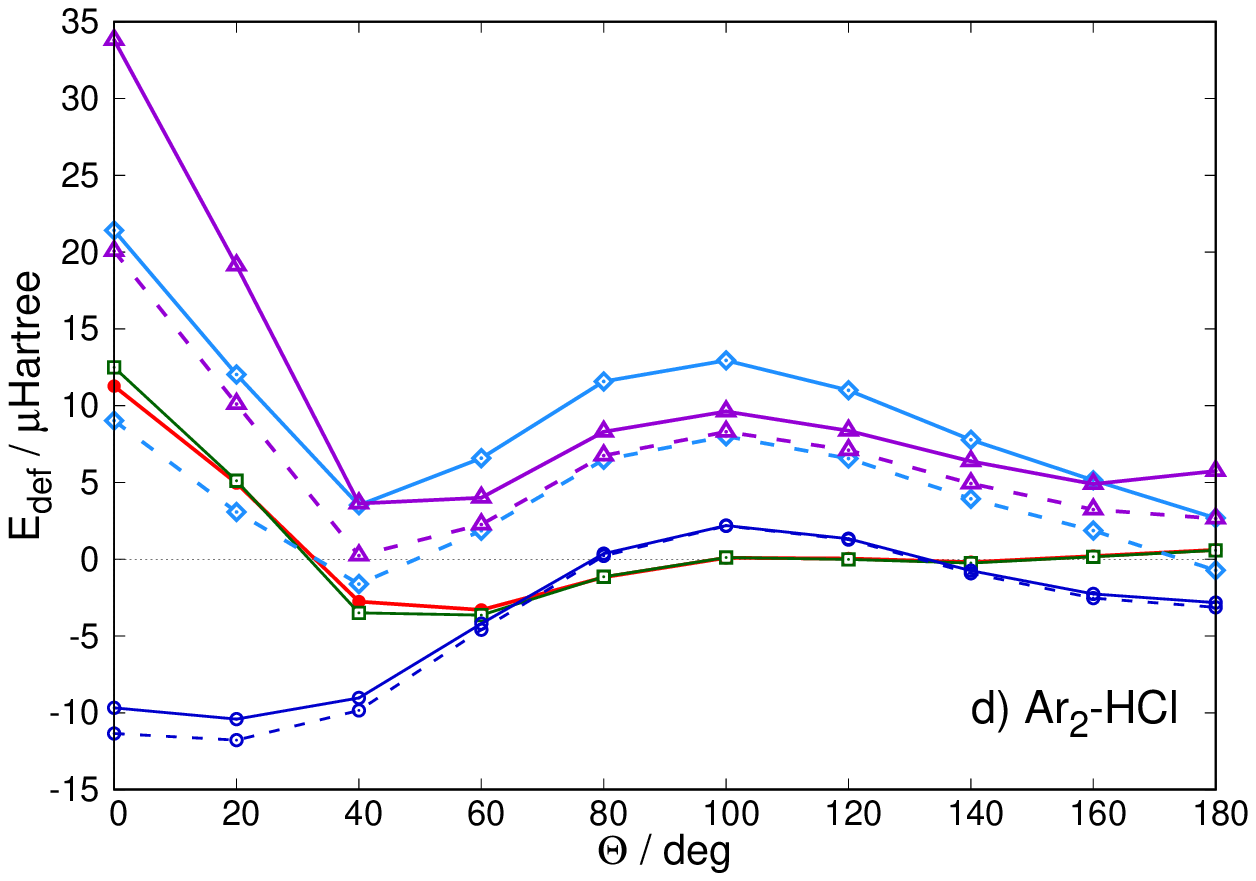} \\
\end{tabular}
\caption{Components  of the 3-body nonadditive interaction energy ($\mu$Hartree) for the in-plane bend of Ar$_2$-HF and Ar$_2$-HCl. Panels $a$ and $c$ show the nonadditive exchange energy. Panels $b$ and $d$ show the deformation energy.
PBdf results are obtained with the PBE0 functional.}
\label{fig:hldef}
\end{figure}

The semilocal (PBE) and global hybrid (PBE0) functionals based on the PBE model heavily overestimate the nonadditive exchange, $E_{\rm nadd-ex}$, with respect to the reference $E^{\rm HL(KS)}_{\rm int}$
results based on DFT-SAPT (Figure~\ref{fig:hldef}). 
As in noble gas trimers, the errors are ameliorated in $\omega$PBE. The PBE-based range-separated functional on the meta-GGA rung, LC-PBETPSS, performs similarly to $\omega$PBE, which confirms that our conclusion on the role of range separated-exchange is general. In addition, the B88-based functionals stay in better agreement with the reference Heitler-London energy than the PBE-based approximations. 

It is tempting to check if going beyond PBE- and B88-type enhancement factors leads to any improvement in the description of the nonadditive exchange. To this end, we examined two recently developed meta-GGA functionals: MVS\cite{Sun:15} and SCAN.\cite{Sun2015b} The enhancement factors built into these formulas have radically different asymptotic behavior from those in PBE and B88.~\cite{Sun:15} The MVS/SCAN enhancement factor satisfies

\begin{equation}
\lim_{s\rightarrow \infty} F_x(s,\alpha) \varpropto s^{-(1/2)},
\end{equation} 
where $\alpha$ is a dimensionless variable which depends on the kinetic energy density.~\cite{Sun:15}
In contrast, the BLYP enhancement factor is approximately linear for large $s$, whereas PBE approaches a constant. 

\begin{figure}
\includegraphics[scale=0.65]{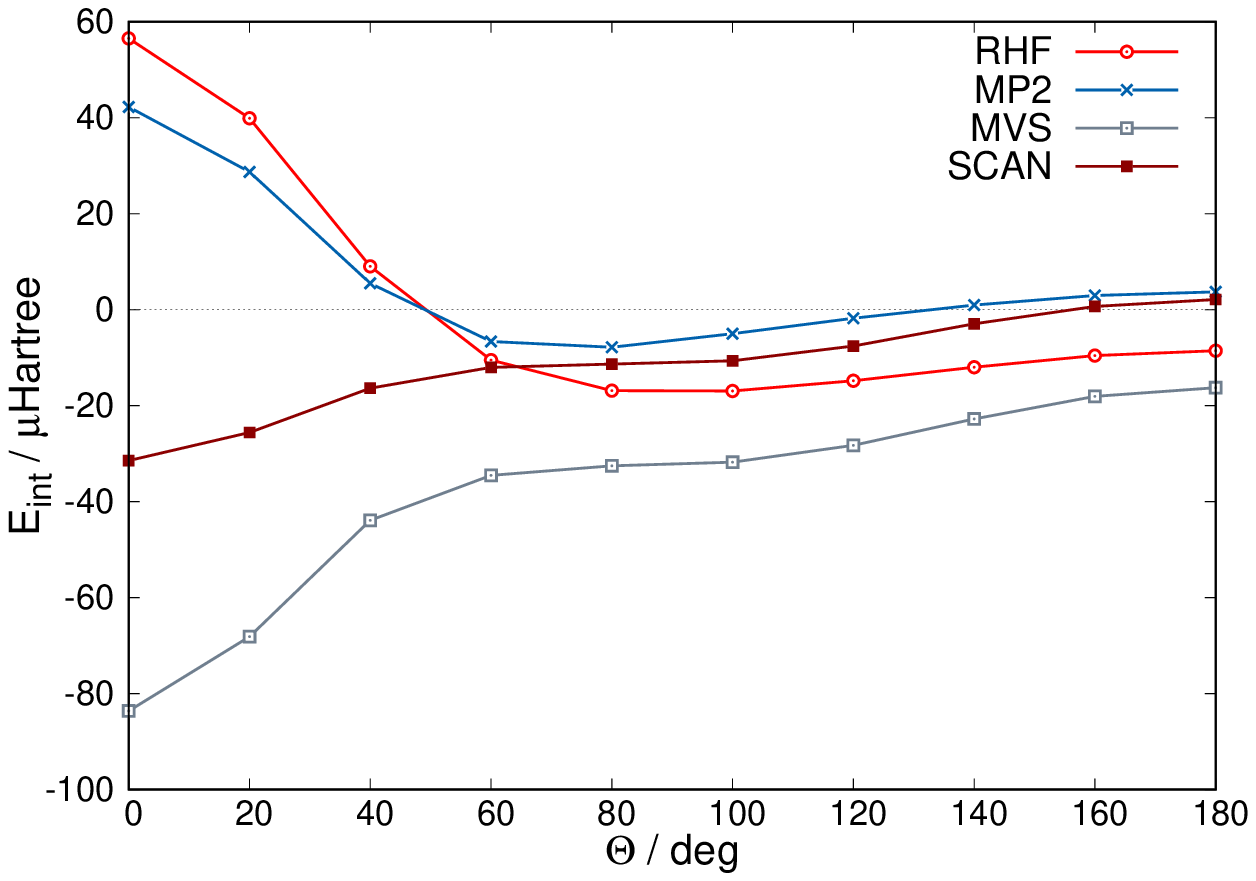} 
\includegraphics[scale=0.65]{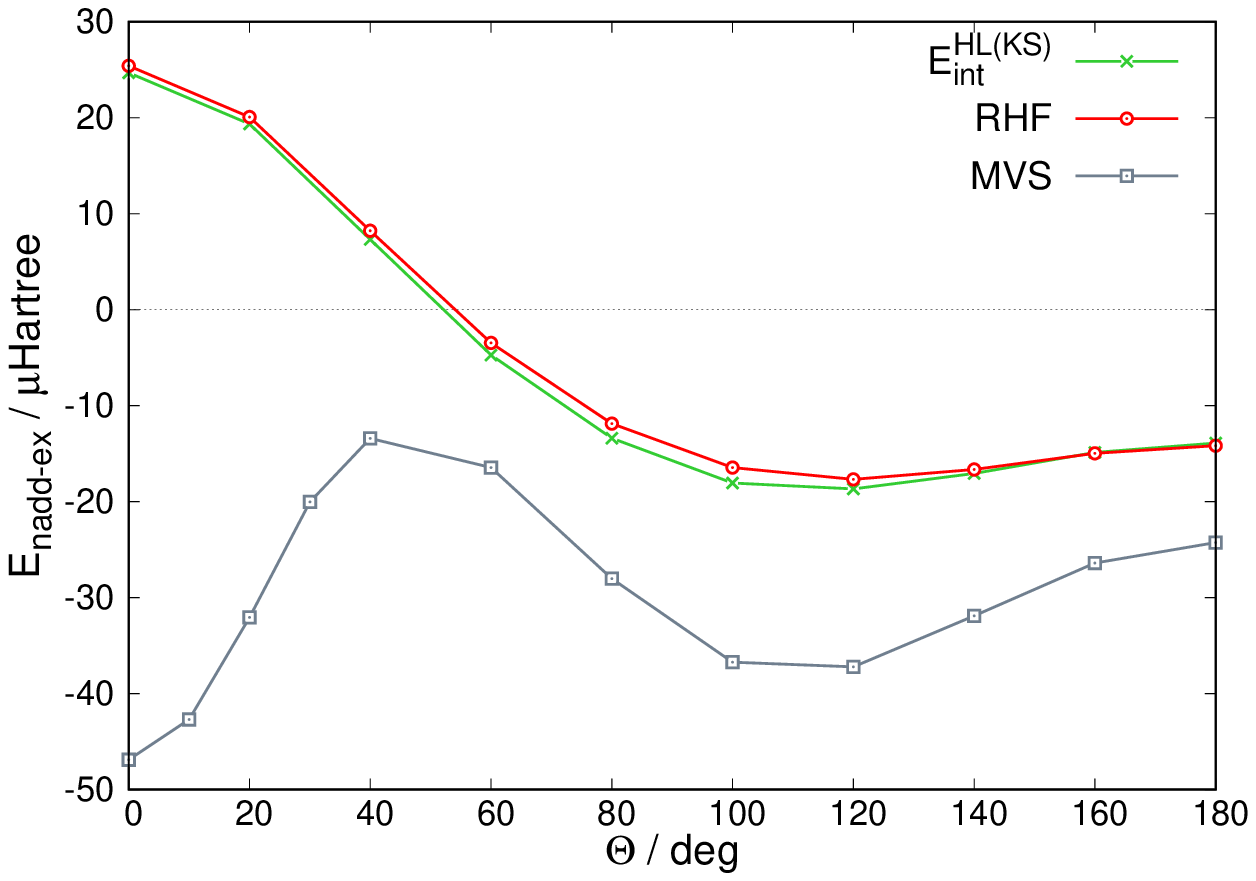} 
\caption{In-plane bend of Ar$_2$-HF. Left: total 3-body nonadditive interaction energy. Right: 3-body nonadditive exchange energy ($\mu$Hartree). }
\label{mvsscan}
\end{figure}

This new $F_x(s)$, however, does not lead to an improvement for 3-body energies in Ar$_2$-HX. For $\Theta \in (\ang{0}, \ang{60})$ both MVS and SCAN predict wrong angular dependence of the total interaction energies (Figure \ref{mvsscan}). The error can be attributed to the behavior of the nonadditive exchange part (for Ar$_2$-HCl results see SI).

The spread of deformation energies computed with different DFT methods is much smaller than in the case of the exchange nonadditivity (Figure~\ref{fig:hldef}). Long range correction no longer moves the results towards Hartree-Fock ones. The curves corresponding to $\omega$PBE and LC-PBETPSS deviate from both Hartree-Fock and global hybrids in the range of $\Theta = \ang{0}$ to \ang{60}.

\begin{figure}
\includegraphics[scale=0.65]{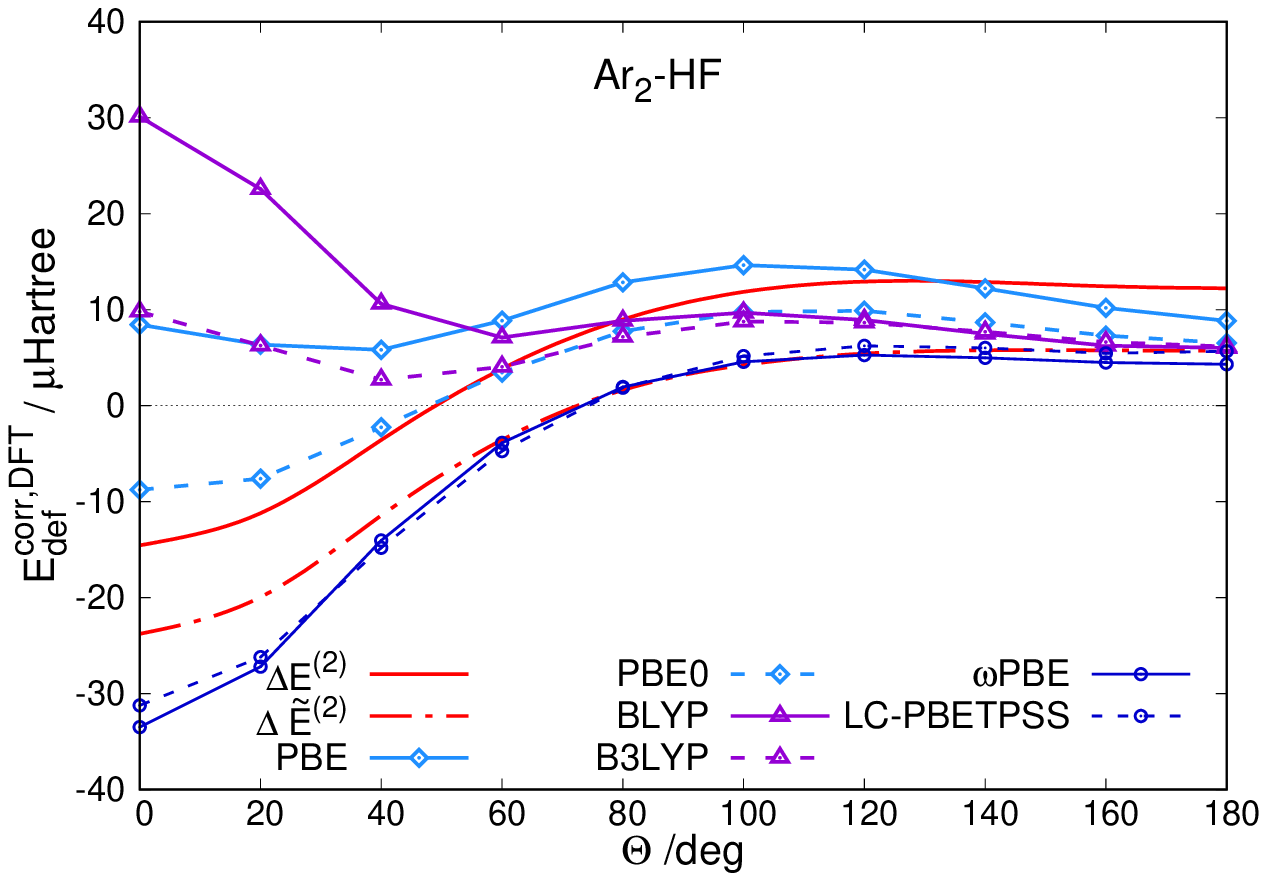} 
\includegraphics[scale=0.65]{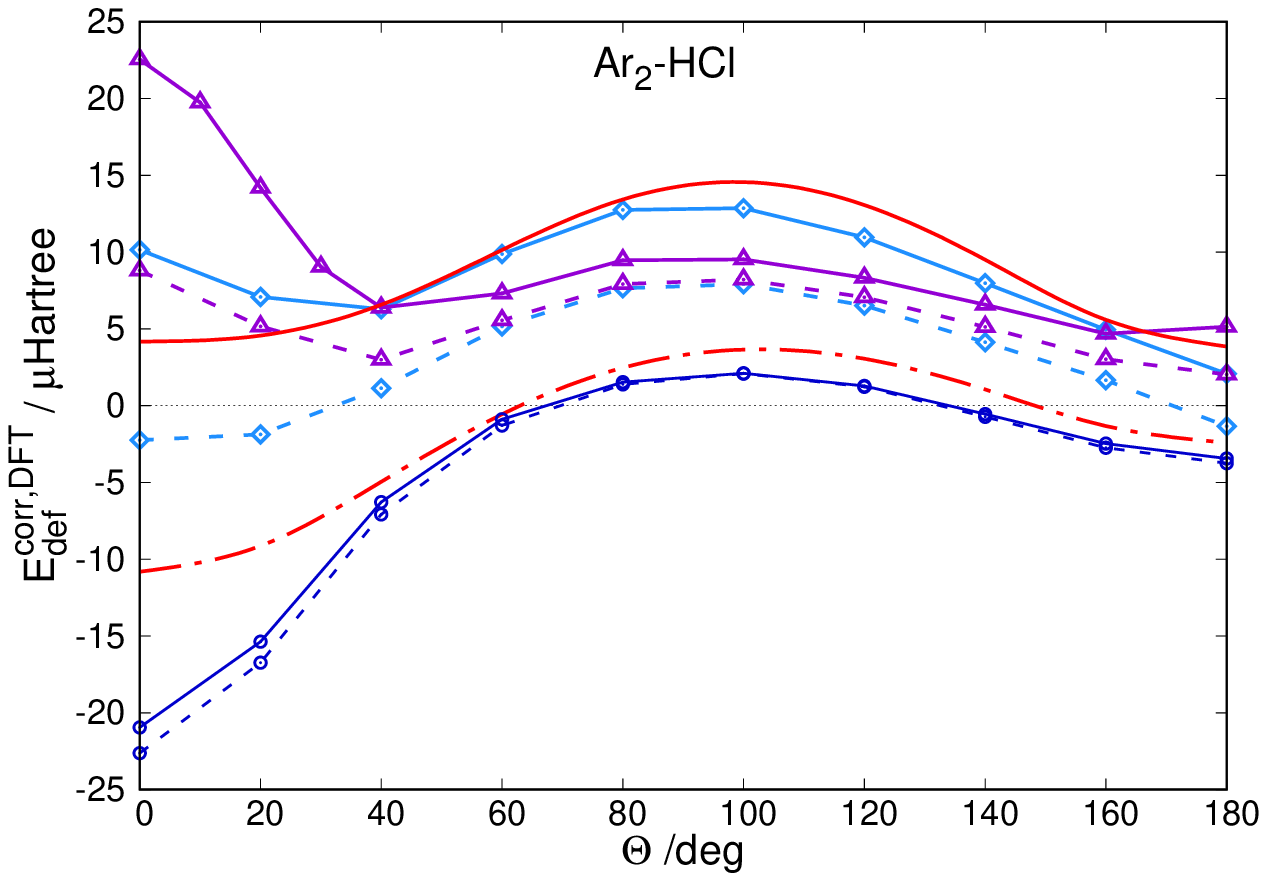} 
\caption{Difference between the 3-body deformation energy at the DFT and Hartree-Fock levels of theory, $E_{\rm def}^{\rm corr,DFT}$, for the in-plane bend of the Ar$_2$-HX trimers. See text for the definitions of $\Delta E^{(2)}$ and~$\Delta \widetilde{E}^{(2)}$.}
\label{fig:indi}
\end{figure}

To understand whether the difference between range-separated DFT and Hartree-Fock is related to any physical effect, we compare
\beq
E_{\rm def}^{\rm corr,DFT} = E_{\rm def}^{\rm DFT}-E_{\rm def}^{\rm HF},
\eeq
with two quantities derived from the MP2 interaction energy:
\beq
\Delta E^{(2)} = E_{\rm int}^{\rm MP2} - E_{\rm int}^{\rm HF}
\eeq
and 
\beq
\Delta \widetilde{E}^{(2)} = E_{\rm int}^{\rm MP2} - E_{\rm int}^{\rm HF} - E^{(2;0)}_{\rm exch-disp}.
\label{eq:indi}
\eeq
$\Delta E^{(2)}$ is the correlation contribution to MP2 interaction energy which contains intramonomer correlation contributions to nonadditive exchange and induction terms, second-order exchange-dispersion, 
and third-order induction-dispersion and exchange-induction-dispersion terms.~\cite{chalbie:90,Chalbie:00} As shown before\cite{Lotrich:1998,Moszynski:1998} the induction-dispersion terms are partly canceled by their exchange counterparts. 

Importantly, the correlation contribution to the nonadditive exchange is small in Ar$_2$-HX. This can be inferred from Figure~\ref{fig:hldef} where the difference between the Heitler-London exchange energy at the DFT-SAPT level of theory $\left(E^{\rm HL(KS)}_{\rm int}\right)$ and Hartree-Fock nonadditive exchange is on the order of 1~$\mu$Hartree (see also Figure~S3 in SI).
Among the physical contributions to $E_{\rm def}^{\rm corr,DFT}$, we expect that the DFT approximations reliably account for the intramonomer correlation contributions to the assorted induction terms. 

The range-separated functionals qualitatively agree with {\mpcorr} while semilocal functionals and global hybrids are close to $\Delta E^{(2)}$ (Figure~\ref{fig:indi}).
This indicates that the former excludes the second-order exchange-dispersion term, while the latter includes it.
The $E^{(2)}_{\rm exch-disp}$ contribution, which is the exchange counterpart of (the additive) second-order dispersion, may appear in semilocal functionals only as an artifact, similar to the fictitious exchange binding of noble gas dimers.
Note that the addition of Hartree-Fock exchange in global hybrids partially removes the spurious angular dependence observed for semilocal exchange at $\Theta < \ang{60}$.

It should be stressed that although the second-order exchange-dispersion is a sizeable effect\cite{Lotrich:1998,Moszynski:1998} (see Table S5), it should not contribute to the damping of the dispersion nonadditivity, the third-order Axilrod-Teller-Muto (ATM) term. This role should be reserved for the exchange counterpart of the third-order ATM dispersion (e.g., Refs.~\onlinecite{Podeszwa07,Podeszwa:2008,Huang:2015}). If $E_{\rm exch-disp}^{(2)}$ proves necessary in the context of DFT, it should be employed in a controllable manner. Therefore, range-separated functionals provide an appropriate starting point for the development of new dispersion-corrected approaches.

\subsection{H-bonds and dispersion in molecular trimers}
We apply our scheme for energy decomposition on a subset of the 3B-69 data set of \v{R}ez\'{a}\v{c} \textit{et al.}~\cite{Rezac:15} (water, formaldehyde, methanol-ethyne, and acetonitrile trimers), three dispersion-bonded (CO, CH$_4$, N$_2$) and two H-bonded (HF, NH$_3$) trimers studied in Ref.~\onlinecite{Erin:2016}. 
Only a subset of 3B-69 data set is included due to our computational limits. Our set includes seventeen configurations in total, see SI for details.

\begin{table}
\caption{Supermolecular ($E_{\rm int}$) and PBdf ($E_{\rm int}^{\rm dfree}$) 3-body nonadditive energies (kcal/mol) for the water trimer (1c geometry from Ref.~\onlinecite{Rezac:15}). The MP2 and CCSD(T) results at the CBS limit are taken from Ref.~\onlinecite{Rezac:15}. All DFT calculations were carried out in the aug-cc-pVQZ basis set.}
\begin{tabular}{l *{6}{S}}
\toprule
{\bf H$_2$O (1c)}   &  \multicolumn{1}{c}{$E_{\rm int}$} & \multicolumn{1}{c}{$E_{\rm int}^{\rm dfree}$} & \multicolumn{1}{c}{$E_{\rm nadd-ex}$} & \multicolumn{1}{c}{$E_{\rm nadd-ex}^{\rm dfree}$} & \multicolumn{1}{c}{$E_{\rm def}$} &  \multicolumn{1}{c}{$E_{\rm def}^{\rm dfree}$}   \\ \midrule
HF   & -2.473 & -2.473 & -0.286 & -0.286 & -2.187 & -2.187  \\
$E_{\rm int}^{\rm HL(KS)}$    & &        & -0.302 & -0.302   &        & \\
PBE  & -2.449 & -2.620 & 0.217  & -0.225 & -2.665 & -2.395 \\
PBE0 & -2.466 & -2.560 & 0.058  & -0.252 & -2.524 & -2.308  \\
BLYP & -2.734 & -2.560 & -0.079 & -0.210 & -2.655 & -2.350 \\
BHLYP & -2.389 & -2.453 & 0.002  & -0.278 & -2.392 & -2.175  \\
MVS   & -2.668 & -2.523 & -0.175 & -0.257 & -2.492 & -2.270 \\
$\omega$PBE & -2.694 & -2.505    & -0.267 & -0.286 & -2.427 & -2.219  \\ 
MP2     &  -2.472 &  &  &  &   & \\  
CCSD(T) &  -2.416 &  &  &  &   & \\ \bottomrule
\end{tabular}
\label{tab:1c}
\end{table}

Different description of the exchange nonadditivity is the major source of variance among DFT approximations (see Tables~\ref{tab:1c} and \ref{tab:ch4}, Tables S7 and S8 in SI). The deformation energy, on the other hand, is much more consistent even between functionals based on different models of the enhancement factor, e.g., PBE and BLYP yield almost identical $E_{\rm def}$ for seventeen tested trimers. 

Remarkably, in contrast to noble gas trimers and Ar$_2$-HX discussed above, in H-bonded systems the exchange nonadditivity is well reproduced by range-separated hybrids. For the water trimer in cyclic configuration $\omega$PBE
gives $E_{\rm nadd-ex} = -0.267$~kcal/mol, which is in reasonable agreement with
the reference $E_{\rm int}^{\rm HL(KS)} = -0.302$~kcal/mol (Table~\ref{tab:1c}). 
Both semilocal and global hybrid approximations predict exchange nonadditivities less accurately than range-separated hybrids.
The MVS functional yields better $E_{\rm nadd-ex}$ than any of the other tested semilocal methods. Moreover, out of seven systems of predominantly H-bonded character, in four cases (HF, \ce{NH3}, methanol-ethyne 3a and 3c) MVS is more accurate than the global hybrids and approaches the accuracy of $\omega$PBE (Tables S7 and S8 in SI).

The remaining error of $\omega$PBE stems mainly from the lack of the exchange-dispersion and dispersion contributions as well as inaccurate deformation due to the residual delocalization error. 
For example, for the water trimer adding the coupled Kohn-Sham $E_{\rm exch-disp}^{(2)} = 0.196$~kcal/mol moves $\omega$PBE close to the MP2 result (Table~\ref{tab:1c}). 
Note that exchange-dispersion and  other correlated contributions at the MP2 level
largely cancel each other in this system, as indicated by the excellent agreement of Hartree-Fock and MP2.

To test how the addition of the exchange-dispersion term affects DFT performance, we added this contribution (at the uncoupled SAPT Hartree-Fock level) to a set of pure, global hybrid and range-separated functionals (Table S9 in SI). As expected from our discussion of Ar$_2$-HX, range-separated functionals benefit the most from the inclusion of $E^{(2)}_{\rm exch-disp}$; the average errors of $\omega$PBE and $\omega$B97XD3 on the set of seventeen trimers are reduced by one-third. 
In contrast, for PBE, PBE0 and BHLYP the average errors become larger. This finding shows that the three-body DFT dispersion correction should depend on the type of the base functional.

As seen in Table~\ref{tab:1c} due to the delocalization error, pure GGA functionals yield inaccurate deformation in H-bonded clusters.
The error diminishes upon admixture of a large portion of exact-exchange (recommended 50\%\cite{Rezac:15,Erin:2016}) or application of long range Hartree-Fock exchange in range-separated hybrids. Accordingly, $\omega$PBE and BHLYP yield similar deformation energies for water trimer: $E_{\rm def}= -2.427$~kcal/mol and -2.392 kcal/mol, respectively.

The good performance of PBE and PBE0 for H-bonded clusters relies on the cancellation of errors between the nonadditive exchange and deformation components (Table~\ref{tab:1c}). This is in accordance with the observation of Ref.~\onlinecite{Erin:2016}. 
One should note, however, that the error cancellation in supermolecular calculations with PBE-based semilocal and global hybrid approximations does not work equally well for every configuration: both PBE and PBE0 typically overestimate the repulsive 3-body energy contributions.~\cite{Rezac:15} On the whole test set dispersion-corrected PBE stands out with its large average error of MAE = 0.093 kcal/mol which is only slightly improved by the introduction of the long range Hartree-Fock exchange in $\omega$PBE (Table~\ref{tab:17}).

The B88-based semilocal and global hybrid approximations improve the nonadditive exchange with respect to the PBE-based methods. 
Still, without the long-range Hartree-Fock exchange it is impossible to realistically describe $E_{\rm nadd-ex}$.
BHLYP supplied with a dispersion correction gives the most accurate total 3-body interaction energies on the whole set of seventeen trimers (Table~\ref{tab:17}). It was also identified as the best-performing functional in the paper of \v{R}ez\'{a}\v{c} \textit{et al.}\cite{Rezac:15} with the root-mean-square error of 0.045 kcal/mol compared to 0.059 kcal/mol for MP2.
However, our energy decomposition has shown that BHLYP relies on error cancellation between the inaccurate exchange nonadditivity and other contributions. For example, if we substituted $E^{\rm HL(KS)}_{\rm int}$ for the BHLYP exchange nonadditivity of the water trimer (Table~\ref{tab:1c}), the total 3-body interaction energy would be the same as that of $\omega$PBE.

The 3-body effects in dispersion-bound trimers (\ce{N2}, CO, and \ce{CH4}) are beyond the reach of DFT approximations even at the qualitative level. None of the DFT functionals correctly accounts for 3-body nonadditive exchange, similarly to noble gas trimers (Table~\ref{tab:ch4}). In contrast, the dispersion-free PBdf approach gives good description of exchange
interactions for both dispersion- and H-bonded clusters. A plausible explanation for the distinction between the DFT treatment
of hydrogen- and dispersion-bonded  systems is
that in the case of the latter the energetic contributions to the interaction energy come from regions of relatively small densities.~\cite{Johnson:2010}

\begin{table}
\caption{Mean absolute errors (kcal/mol) with respect to $E_{\rm int}$(CCSD(T)) for the set of 17 trimers. The supermolecular Hartree-Fock and DFT results include D3 dispersion 3-body term based on a damped ATM formula (denoted $E_{\rm disp}^{\rm D3}$).~\cite{Grimme:2010}}
\begin{tabular}{l *{7}{S}}
\toprule
     & \multicolumn{1}{c}{HF}  & \multicolumn{1}{c}{BLYP} & \multicolumn{1}{c}{BHLYP} & \multicolumn{1}{c}{PBE}  & \multicolumn{1}{c}{PBE0} & \multicolumn{1}{c}{$\omega$PBE} & \multicolumn{1}{c}{MP2}   \\ \midrule
$E_{\rm int}+E_{\rm disp}^{\rm D3}$  & 0.032 & 0.108  & 0.032   & 0.093  & 0.058 & 0.079 & 0.020 \\
$E_{\rm int}^{\rm dfree}+E_{\rm disp}^{\rm D3}$ & 0.032 & 0.046  & 0.024   & 0.063  & 0.044 & 0.031 &  \\
\bottomrule
\end{tabular}
\label{tab:17}
\end{table}

\begin{table}[h!]
\caption{Supermolecular ($E_{\rm int}$) and PBdf ($E_{\rm int}^{\rm dfree}$) 3-body nonadditive energies ($\mu$Hartree) for the CH$_4$ trimer. Geometry taken from Ref.~\onlinecite{Erin:2016}. All calculations were performed in the aug-cc-pVTZ basis set.}
\begin{tabular}{l c p{2mm} c p{2mm} c p{3mm} c p{2mm} c p{2mm} c }
\toprule
{\bf \ce{CH4}}  & $E_{\rm int}$ &&  $E_{\rm int}^{\rm dfree}$ && $E_{\rm nadd-ex}$ && $E_{\rm nadd-ex}^{\rm dfree}$ && $E_{\rm def}$ &&  $E_{\rm def}^{\rm dfree}$ \\ \hline
HF      & -21.22 && -21.22 && -20.40 && -20.40 && -0.819 && -0.819  \\
$E^{\rm HL(KS)}_{\rm int}$ &     &&  && -34.24 && -34.24 &&  &&  \\
PBE     & 187.4  && -27.31 && 176.5  && -26.40 && 10.84  && -0.908 \\
PBE0    & 105.8  && -25.32 && 101.8  && -24.17 && 3.976  && -1.149 \\
BLYP    & -57.51 && -22.48 && -69.24 && -21.94 && 11.73  && -0.534 \\
BHLYP   & -5.155 && -22.35 && -8.696 && -21.61 && 3.541  && -0.740 \\
MVS     & -204.0 && -21.77 && -187.7 && -20.13 && -16.22 && -1.639 \\
$\omega$PBE & -80.23 && -26.53 && -70.16 && -24.63 && -10.07 && -1.900  \\
MP2     & -5.241  &&        &&        &&        && &&  \\  
CCSD(T) & 24.66   &&        &&        &&        && &&  \\ \hline
\end{tabular}
\label{tab:ch4}
\end{table}

\section{Conclusions}
This work has examined the components of nonadditive interactions obtained in the supermolecular DFT calculations.
We proposed a physically meaningful decompos​ition​ of the total 3-body​ nonadditive​ DFT interaction energy into the exchange interaction term, which captures the effect of the Pauli exclusion in the wave function of the trimer, and the deformation energy, which originates from the mutual polarization of the monomers. The scheme, known as Pauli Blockade, appears to be an essential tool for understanding the performance and future improvements of DFT models for noncovalent many-body systems.

The major source of variance among DFT methods is the nonadditive exchange term. The behavior of this contribution can be discussed for two limiting cases: dispersion and hydrogen-bonded  clusters. 
For the former no DFT approximation reproduces nonadditive exchange even at a qualitative level, as shown e.g., for rare gas trimers, Ar$_2$-HX and methane trimer. For H-bonded systems range-separated functionals are the only type of DFT approximations capable of describing the nonadditive exchange (to within 20\% of the reference value). For the paradigm Ar$_2$-HX (X=F,Cl) systems, range-separated hybrids yield correct anisotropy of $E_{\rm nadd-ex}$,​ but​ the interaction curves​ are significantly shifted from the reference value​s​. This pertains to both GGA and meta-GGA range-separated hybrids, as nonempirical functionals belonging to both rungs, i.e. $\omega$PBE and LC-PBETPSS, respectively, are tested.
The MVS and SCAN functionals employing the latest ideas in the enhancement factor design fail in these trimers.

The DFT errors in the deformation energy are smaller than those of the nonadditive exchange, and the differences between functionals are less pronounced.
The correlation effects on induction terms could be indirectly probed in specific instances such as Ar$_2$-HX. We showed that range-separated functionals can recover these effects semiquantitatively. Other DFA types do not permit a more detailed interpretation because their deformation effects are obscured by artifacts of the exchange functionals.

The second order exchange-dispersion nonadditivity was shown to be very important in the wide range of systems.~\cite{Podeszwa07}
This nonadditivity represents the exchange counterpart of the additive second-order dispersion term and is completely ignored in the existing Axilrod-Teller-Muto force-field-like dispersion corrections, due to a completely different functional dependence. In the majority of studied situations, the second-order exchange-dispersion either exceeds or nearly equals the third-order ATM nonadditivity. For example, in the cyclic water trimer (1c) the coupled Kohn-Sham $E^{(2)}_{\rm exch-disp}$ equals 0.196 kcal/mol whereas 
an approximate damped ATM term reported in Ref.~\onlinecite{Rezac:15} amounts to 0.028 kcal/mol.
Our results for Ar$_2$-HX suggest that range-separated hybrids do not account for $E^{(2)}_{\rm exch-disp}$, while semilocal and global hybrid functionals may imitate this contribution in the short range. Consequently, on a larger set of molecules, supplementing the existing dispersion corrections with $E^{(2)}_{\rm exch-disp}$ term considerably improves the error statistics for range-separated functionals, but not for pure and global hybrid approximations.

The dispersion-free Pauli Blockade​ scheme​, in which the mutual polarization​ ​of the monomers occurs via HF operators, fully corrects the description of the nonadditive exchange energy for every functional and every system, in particular, for both hydrogen- and dispersion-bonded trimers. 
By comparison with DFT-SAPT we have shown that PBdf can be used as reference for the assessment of nonadditive exchange. It should be noted that in PBdf the nonadditive exchange-dispersion term is not accounted for.

The former can be ameliorated in the particular case of hydrogen-bonded systems by the use of exact exchange at long range.
To conclude, in designing DFT approximations for description of nonadditive 3-body interactions in weakly bound systems, the major challenges are related to a reliable modeling of the nonadditivities of the first-order as well as the exchange-dispersion effects.
The latter presents a challenge for all types of systems and warrants going beyond the expressions currently used for three-body dispersion corrections.​ 

\section{Supplementary Material}
Supplementary material includes the derivation of the $\Delta_{\rm M}$ term and PB(df) methods, sensitivity test of the $E_{\rm int}^{\rm HL(KS)}$ term, and detailed numerical data for systems presented in Section III.

\section{Acknowledgments}
M. H. and M. M were supported by the Polish Ministry of Science and Higher Education (Grants No. 2014/15/N/ST4/02179 and No. 2014/15/N/ST4/02170). This work was partly supported by the National Science Foundation (Grant CHE-1152474). 

%

\end{document}